\documentclass[showkeys,aps,pre,reprint,superscriptaddress,showpacs,preprintnumbers,amsmath,amssymb,nofootinbib]{revtex4-2}
\usepackage{graphicx}
\usepackage{dcolumn}
\usepackage{bm}

\begin{document}

\preprint{}

\title{Dynamical cooperation model \\ for mitigating the segregation phase in Schelling's model}

\author{Akihisa Okada}
\email{a-okada@mosk.tytlabs.co.jp}

\author{Daisuke Inoue}
\author{Shihori Koyama}
\author{Tadayoshi Matsumori}
\author{Hiroaki Yoshida}

\affiliation{%
Applied Mathematics Research Domain, Toyota Central R\&D Labs., Inc., \\ 1-4-14 Koraku, Bunkyo-ku, Tokyo, 112-0004, Japan
}%

\date{\today}

\begin{abstract}
We consider a Schelling-like segregation model, in which the behavior of individual agents is determined by a mixed individual and global utility. With a high ratio of global utility being incorporated, the agents are cooperative in order to realize a homogenized state, otherwise the agents are less cooperative, leading to an undesired Nash equilibrium with low utility. In the present study, we introduce a dynamically varying cooperation degree parameter to prevent the agents from falling into such a low-utility equilibrium state. More precisely, a large cooperation degree is assigned when the agents are in high-utility regions, whereas agents having low utility behave more individually. Simulation results show that homogenized phases with globally high utility are achieved with the present dynamical control, even for the case of a low mean value of cooperation degree. Since the cooperation degree represents {the magnitude with which Pigouvian tax is enforced} in the model of residential movement within a city, this result suggests the possibility of tax intervention to circumvent the undesired segregation of residents.
\end{abstract}

\keywords{Sociophysics, Schelling model, Nash equilibrium, Segregation, Agent model}
\maketitle


\section{Introduction}
The emergence of collective behavior originating from simple individual constituents is observed in many fields, including not only the classical examples of phase transitions of the system consisting of atoms~\cite{landau1980ii,AGD,Rossler2004,hansen2013},
but also the consciousness arising in neuron networks~\cite{changeux2009physiology} and the self-organized behavior of animal swarms~\cite{bonabeau1999swarm,camazine2003self,shoji2014localized}.
The segregation of residents or urban centralization in social science is attracting increasing interest, and this indeed results from intricate relations between the macroscopic perspectives and the microscopic individual agents.   
In a seminal study, Schelling proposed a multi-agent-based model explaining the resident segregation occurring 
as a consequence of individual preferences for neighborhood environments~\cite{Schelling1971}. 
The degree of satisfaction of each agent for being together with agents of the same group is measured by an individual utility function, and the agent continuously changes location such that the individual utility increases. 
Such a microscopic movement of agents with a relatively small preference for neighbors leads to a macroscopic segregation in which the individual utility is not necessarily maximized. 
This paradoxical result has attracted attention, and this model has been extended to various situations other than the segregation of two race groups, 
such as the problem of population density distribution in a city introducing corresponding utility functions ~\cite{Pancs2007,Clark2008,GRAUWIN2012,Hatna2012}, 
developing statistical physics methods to analyze the equilibrium state~\cite{Vinkovic2006,Stauffer2007,Gauvin2009, Grauwin2009}, 
game theory methods~\cite{Zhang2004,Zhang2011},
extension of the definition of a neighborhood in the model~\cite{Laurie2003},
and linking to a wealth state~\cite{Saha2016}.  
Interestingly, interest in Schelling's research has been renewed rather recently for its analogy with the physicochemical phase transition phenomena, and the corresponding study has been cited from many different aspects, including physics~\cite{Vinkovic2006, PhysRevLett.113.088701} and computer science~\cite{de2009optimal,izquierdo,CORTEZ201560}. 

An illustrative comparison between the effects of collective and individual behaviors 
was made by Grauwin \textit{et al.}~\cite{Grauwin2009}, 
where the model considers a utility function for each agent that takes into account 
both individual preference and the global happiness of the entire city. 
More precisely, the utility function of an agent is constructed as a linear combination of 
the individual utility of the agent and the sum of the utility values over all agents in the city. 
The coupling coefficient thus indicates the degree of cooperation.
A high ratio of global utility indicates that the agents cooperatively 
consider global happiness, and a low ratio indicates that agents are selfish, leading to an undesired Nash equilibrium~\cite{Nash1950} with low utility, as mentioned in the previous paragraph. 
Since the cooperation degree is regarded as {a parameter representing the magnitude of } the Pigouvian tax of a simplified social model of 
residential movement, a high tax is required in order to avoid such a low-utility Nash equilibrium~\cite{Grauwin2009,WANG2015388}. 
{Then the idea of taxing only where necessary is more readily accepted by residents, rather than a situation of uniformly heavy taxation, and it is thus important to examine the effects the varying cooperation degree.}
A result that is useful for this issue was reported by Jensen \textit{et al.}~\cite{Jensen2018}, 
where a relatively small number of individuals behaving altruistically has the power 
to change the equilibrium state, implying that an extra tax on these individuals could significantly 
raise the system utility and break the undesired Nash equilibrium. 
However, it is still desirable that the Nash equilibrium be relaxed with the same degree of cooperation, i.e., 
with no extra tax, although studies on such an active control of the phase transition remain rare. 

In the present study, we consider a dynamically varying cooperation degree parameter aimed at preventing agents from falling into an undesired equilibrium state. 
The rule governing the dynamical variation of the parameter is designed 
such that agents are relatively cooperative when they have high utility, 
and, in contrast, agents having low utility can behave individually. 
We perform numerical analysis of the proposed dynamical model in order to compare the results with 
those obtained for the case of a fixed cooperative degree parameter, 
while maintaining the mean value of the parameter is common for the dynamical and static cases. 
The results show that unsegregated states with globally high utility are achieved with the present dynamical control, even if the corresponding static case with the same mean value of the cooperative degree parameter falls into a segregated equilibrium state. 
These results possibly pave the way for tax intervention to circumvent the undesired segregation of residents.

\section{Model} \label{sec:model}
We first introduce a model describing the motion of agents, in which the parameter for the cooperation degree is important for the decision making of the agents. 
We then present a rule for dynamical variation of the cooperation degree parameter,
which is designed to avoid undesired equilibrium states.

\subsection{Schelling-like segregation model} \label{subsec:prior}
\begin{figure}
	\centering
	\includegraphics[height=4cm]{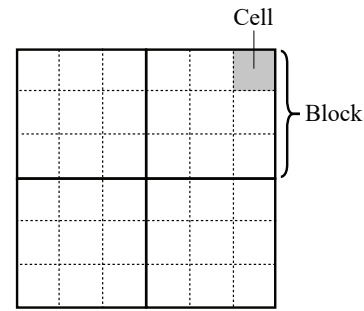}
	\caption{{Illustration of a city with $Q=4$ and $H=9$. The solid lines and the dashed lines represent boundaries between blocks and cells, respectively.}}
	\label{fig:town}
\end{figure}

Let a cell be a place where one person (or family) lives, and a block be a group of cells representing the neighborhood of these people. The entire system representing a city consists of a collection of blocks.
The number of cells in a block and the number of blocks in the city are denoted by $H$ and $Q$, respectively.
Here, we assume that all of the blocks have a common number of cells.
The state of a cell is either vacant or occupied by a resident, which is expressed as a binary variable $x_i^q$ with $i=1,\ldots,H$,
and $q=1,\ldots,Q$, which are indexes for cells and blocks, respectively.
The value of $x_i^q$ is unity if the $i$th cell of $q$th block is occupied and is zero otherwise.
The number of residents $N$ is fixed ($N=\sum_{q=1}^{Q} \sum_{i=1}^{H} x_i^q$),
and accordingly the density of residents in the city is constant ($\rho_0 = N / QH$).
Figure~\ref{fig:town} illustrates an example city with $Q=4$ and $H=9$. 

All residents in a block share a common utility function $u(\rho_q)$, where $\rho_q$ is the density of residents in the $q$th block and is defined as $\rho_q=1/H \sum_{i=1}^H x_i^q$. 
The higher the value of the utility function becomes, the happier the occupants feel and tend to remain in place. 
The sum of the utilities of all residents $U = H \sum_{q=1}^{Q} \rho_q u(\rho_q)$ is thus the utility of the entire city.
The residents move in order to increase the utilities, and the population distribution changes.
Since the movement of an individual affects all of the residents in the source and destination blocks, the change in utility of individuals who have moved $\Delta u$ is generally different from the change in the social utility $\Delta U$. 
Depending on the priority of these two utilities, the equilibrium state toward which the city tends will be different.
In order to quantify the priority for the utilities, a parameter $\alpha$ is introduced as described below, following the model in a previous study~\cite{Grauwin2009}.
The change in utility $\mathcal{G}$ that would be caused by the move of an agent from the $q_1$th block to the $q_2$th block is given by
\begin{equation}
	\mathcal{G} = \Delta u + \alpha \left( \Delta U - \Delta u \right),
\end{equation}
where $\Delta u= u(\rho{^\prime}_{q2})-u(\rho_{q1})$, and $\Delta U = H(\rho'_{q2}u(\rho'_{q2})-\rho_{q2}u(\rho_{q2}))+
H(\rho'_{q1}u(\rho'_{q1})-\rho_{q1}u(\rho_{q1})$, with $\rho'_{q}$ being the density of the $q$th block after the move.
Then, the actual move of the agent from the $q_1$th block to the $q_2$th block takes place with the probability defined as $P = 1 / (1+e^{-\mathcal{G}/T})$.
Here, $T$ is a parameter controlling the fluctuation of the decision.
Since the value of $\alpha$ is the ratio of $\Delta U$, i.e., the utility change of the entire city, 
the value of $\alpha$ is interpreted as a parameter representing {the magnitude of }the Pigouvian tax.
In the previous study \cite{Grauwin2009}, it was shown that for a tax $\alpha=0$, i.e., agents move considering only their own utility, the system tended to fall into an undesired Nash equilibrium state with segregation, which could be avoided with increasing $\alpha$.

\begin{figure}[t]
	\centering
	\includegraphics[width=8cm]{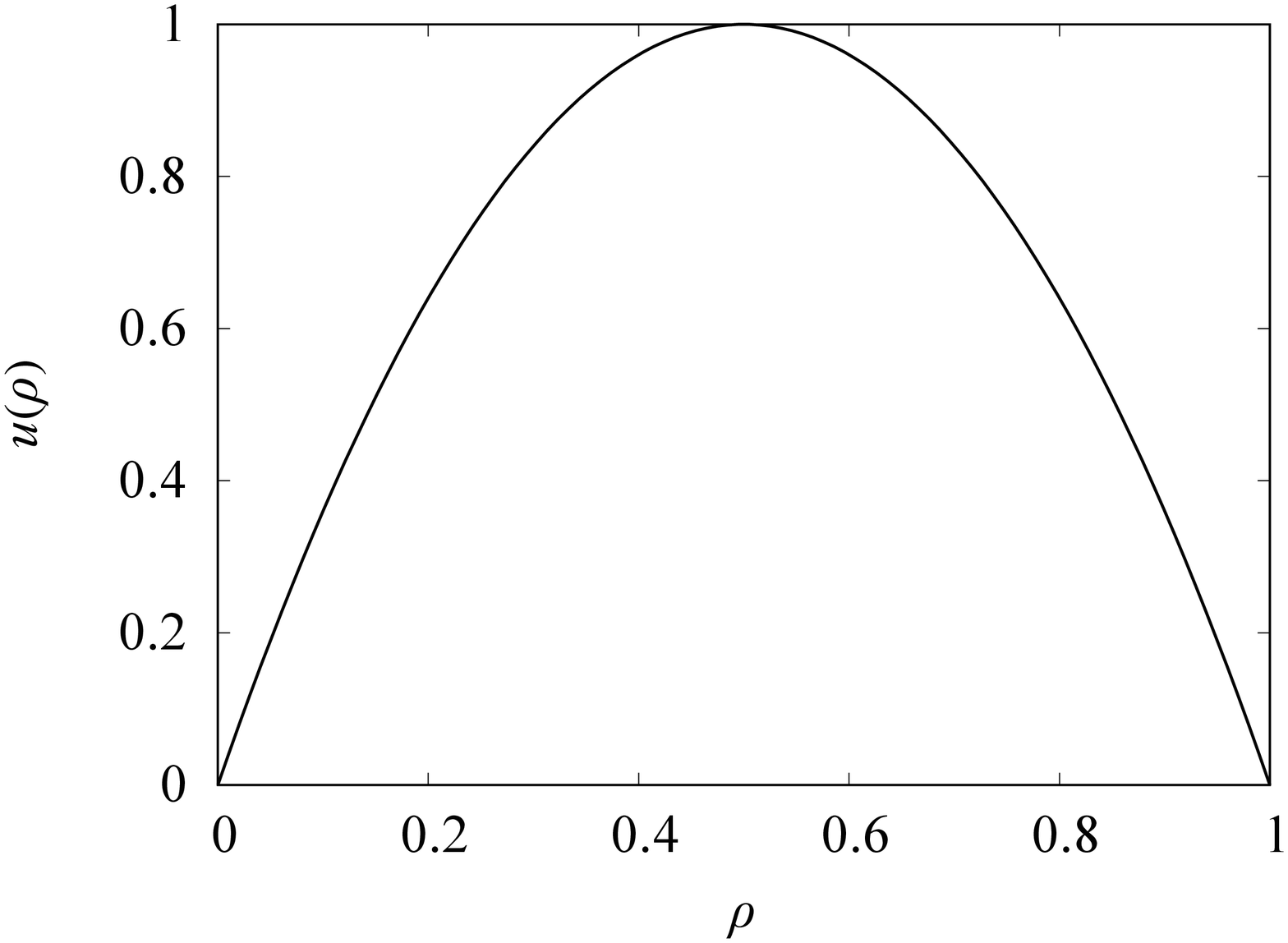}
	\caption{Utility function $u$ given as $u(\rho) = 4 \rho (1 - \rho) $ for a block density $\rho$.}
	\label{fig:utility_func}
\end{figure}

Postulating that the people are happy with an appropriate number of neighbors, and that
they prefer to be neither isolated nor overcrowded, the utility as a function of density is commonly convex upward~
\cite{Pancs2007, Grauwin2009, FLAIG2019}.
In the present study, the individual utility function is defined as quadratic function $u(\rho) = 4 \rho (1 - \rho ) $, as shown in Fig. \ref{fig:utility_func}.
In other words, an agent has the highest utility at $\rho=0.5$ and has zero utility at $\rho=0$ and $1$.

\subsection{Dynamically varying cooperation degree}\label{subsec:dynamic}
In previous studies concerning the Schelling-like segregation model, the parameter for the cooperation degree $\alpha$ is constant in space and time.
If we are interested in the equilibrium state of a city with a fixed policy, the setup with a constant $\alpha$ is sufficient.
In order to actively control the behavior of the agents, however, a policy that takes into account the current situation  is required.
We therefore introduce a dynamical and location-dependent $\alpha$, aimed at realizing a high-utility state for a given mean value of $\alpha$.
After describing the dynamics of movement of the agents and the timing of changing the value of $\alpha$,
we introduce a form of $\alpha$ as a function of $\rho_q$, resulting in different values of tax depending on the situation.
{Here, we use the term ``dynamic'' to mean that the tax changes after every period of time, i.e., it has an implicit time dependence via the time-dependent density.}

In previous studies, e.g., Refs.~\cite{GRAUWIN2012, Grauwin2009, Jensen2018}, 
an equilibrium state with a repeating and sufficient number of trials with the above-described probability is sought.
In each trial, an agent candidate who would move and a vacant cell are selected randomly.
This quasi-dynamical process corresponds to searching for an optimal configuration for all the agents in the city, and every agent has the opportunity to move.
On the other hand, in reality, agents who move at a certain time, such as at the beginning of a semester, could be a certain ratio of the agents.
If we are to consider a rule for the dynamical variation of $\alpha$, the latter situation should be taken into account.
We therefore introduce an interval in which only some agents have the opportunity to move, and
look for an optimal configuration at each interval by repeating the trials.
Here, the ratio of agents who could move during one interval to all agents in the city is denoted by $r$.
Then, between the intervals, we dynamically update the value of $\alpha$ according to a predefined rule, while taking into account the configuration (or the density distribution) of the agents. In other words, this idea focuses on the non-equilibrium aspect of the formation of the city configuration by repeating partial optimizations.

\begin{figure*}[t]
	\centering
	\includegraphics[width=14cm]{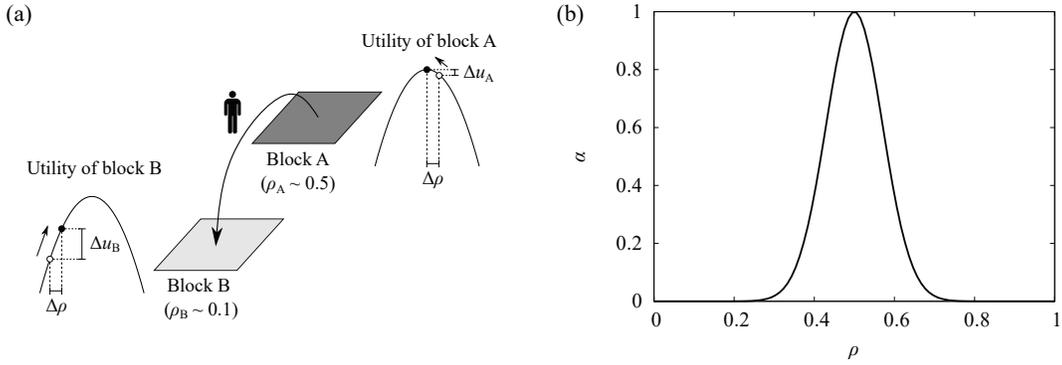}
	\caption{Defined dynamical tax. (a) Schematic diagram of utility variation. If $\alpha \sim 1$ for blocks of $\rho \sim 0.5$, {then} agents in block A could move to block B (lower density: $\rho \sim 0.1$) in order to increase the utilities of neighborhoods. A detailed explanation is given in the main text. (b) Density dependence of cooperation degree $\alpha$, which is expressed as a Gaussian function $\exp\left[ - ( \rho - 0.5)^2 / 2\sigma^2 \right]$ of block density $\rho$.}  
	\label{fig:tax_func}
\end{figure*}

Next, we present a rule to vary the value of $\alpha$ dynamically between $0$ and $1$.
We first describe a typical situation in which an undesired Nash equilibrium state is established.
For simplicity, we consider the situation shown in Fig.~\ref{fig:tax_func}(a). 
The density of block A is $\rho \sim 0.5$, and the density of block B is $\rho \sim 0.1$.
This means that the utility of residents in block A is higher than the utility of residents in block B.
We consider whether the move of an agent is adopted for the case in which block A and block B are selected as a source and destination, respectively.
Obviously, the individual utility of the agent is decreased, and therefore this type of move is declined with the low value of $\alpha$.
This rejection of moving could happen even if the density of block A is slightly larger (and thus a lower utility of agents in block A), which can cause an undesired Nash equilibrium with low total utility. 
On the other hand, the change of density in block B in Fig.~\ref{fig:tax_func}(a) results in utility gains of the other agents in block B. 
Thus, for a high value of $\alpha$, this type of move is accepted if
the sum of these utility gains of other agents in block B exceeds the utility loss of the individual utility of the mover.
This altruistic behavior is the source of elimination of the undesired Nash equilibrium, as pointed out in the previous study~\cite{Grauwin2009} with static cooperation degree parameter $\alpha$.

Keeping in mind the importance of moving from a good density region, we design the dynamical cooperation degree parameter $\alpha$
such that the value of $\alpha$ is high around $\rho=0.5$ and low otherwise.
Specifically, we use the following Gaussian function, which is already normalized as $0 \le \alpha \le 1$:
\begin{equation}
	\alpha(\rho_i) = \exp \left[ -\frac{\left( \rho_i - 0.5 \right)^2}{2\sigma^2} \right],
\end{equation}
where $\sigma$ represents the width of the high-$\alpha$ region.
{We employ this form as a representative expression for the convenience of controlling the shape via the variance,
	though any function taking its maximum value at the center, such as a triangle, could work in the same manner.}
The case of $\sigma=0.1$ is shown in Fig. \ref{fig:tax_func}(b).

{We qualitatively explain the reason why this shape function of $\alpha$, in which high value is assigned to high utility residents, could prevent segregation. A resident in a $\rho \sim 0.5$ block, which means the resident has high $\alpha$ in our setting, can move to a block with lower density, considering the utility of the entire city. On the other hand, residents in a low density state do not need to have a high cooperativity $\alpha$ because their behavior originally matches the increase in individual utility with the increase in citywide utility.}

The value of $\sigma$ is determined such that the block average of $\alpha$ is equal to a specified value:
\begin{equation}
	\bar{\alpha} \equiv \frac{1}{H} \sum_{i=1}^{H} \alpha(\rho_i) = \alpha_\textrm{s}.
\end{equation}
{Here we remark on the choice of the block average rather than the agent average.  Reference~\cite{Grauwin2009} showed that the states resulting from iterative computations of individual moves statistically coincide with equilibrium states.
	In this respect, a random movement of an agent is merely a virtual movement performed to find the state of optimal choice in the equilibrium state. Therefore it is natural to set $\alpha$ for the state of each block, rather than setting $\alpha$ each time for the individuals.}

In the present study, comparison with the previous static case is made with the static $\alpha$ being the same as this average value $\alpha_\textrm{s}$.
{For the case of $\alpha_\textrm{s}=0$ or $\alpha_\textrm{s}=1$, we directly substitute the values instead of adjusting $\sigma\to 0$ or $\sigma\to \infty$.}

\begin{figure*}[t]
	\centering
	\includegraphics[width=10cm]{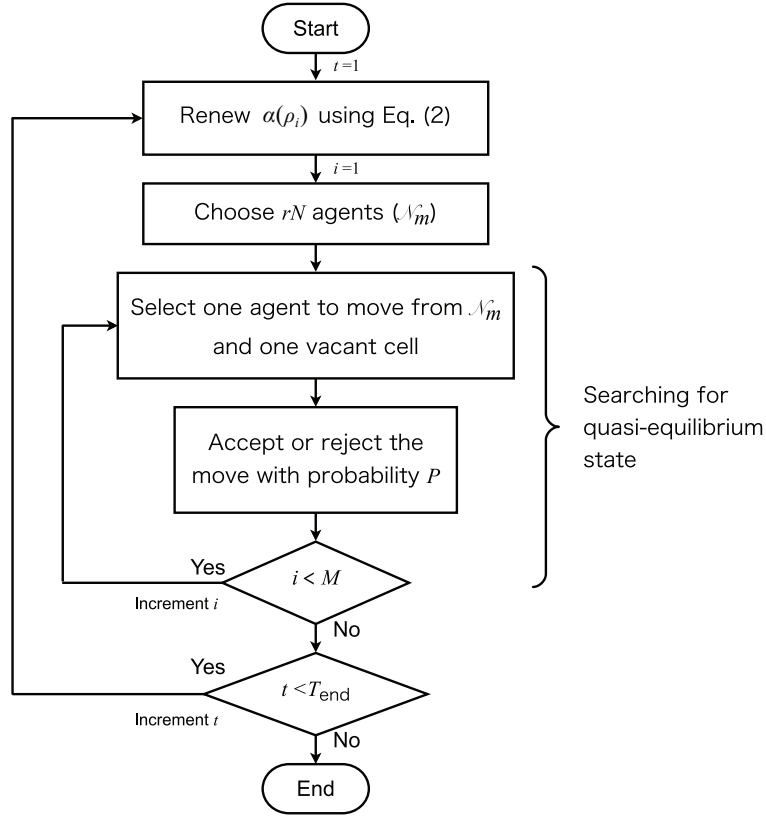}
	\caption{Flowchart of the actual implementation of the dynamical cooperation degree model.
		The parameters $N$, $M$, and $T_\textrm{end}$ are the total number of agents, 
		the number of trials to reach the quasi-equilibrium state in one interval, 
		and the number of time steps to reach the real equilibrium state of the city, respectively.}
	\label{fig:flowchart}  
\end{figure*}

\subsection{Procedure and parameter values}
In our algorithm, the quasi-equilibrium state with $rN$ moving agents is searched for at each time interval, with redistributed values of cooperation degree $\alpha$.
The equilibrium state of the entire city is reached by repeating this searching process.
The time dependence of the utility per agent is examined in terms of
the interval for searching for the quasi-equilibrium state as a unit time.
We summarize the actual implementation of our dynamical cooperation degree model
as a flowchart in Fig. \ref{fig:flowchart}.
In each interval in which rate $r$ of the agents move, $M=50\,000$ trials are made in order to reach the quasi-equilibrium state at the moment,
{the validity of which will be confirmed in the following section.}
One process of this interval corresponds to one time step, which is repeated $T_{\mathrm{end}}$ times while varying $\alpha$.
The value of $T_{\mathrm{end}}$ is chosen such that the system reaches the real equilibrium state, 
typically four times the inverse of $r$, e.g., for $r=0.1$, the process of reaching the quasi-equilibrium state is repeated $40$ times.
The parameters used in the analysis are summarized in Table~\ref{table:parameter}. 
The initial placement of agents within the city is uniformly randomized.
Since the present model includes a stochastic process, in general we perform numerical computations multiple times, typically three times, and the mean value is plotted with variance as error bars. 

\begin{table*}
	\centering
	\begin{minipage}{0.4\linewidth	}
		\centering
		\caption{Parameters used in analysis.}
		\label{table:parameter}
		\begin{tabular}{c|l}
			\hline
			Variable & Value\\ \hline
			$Q$ & 25 \\
			$H$ & 100 \\
			$T$ & 0.01 \\
			$\bar{\alpha}$ & 0 to 1 in increments of {0.1} \\
			$\rho_0$ & 0.1 to 0.5 in increments of 0.1 \\ \hline
		\end{tabular}
	\end{minipage}
\end{table*}

\section{Results and discussion}

\subsection{Per-agent utility values}
\begin{figure}[t]
	\centering
	\includegraphics[width=8cm]{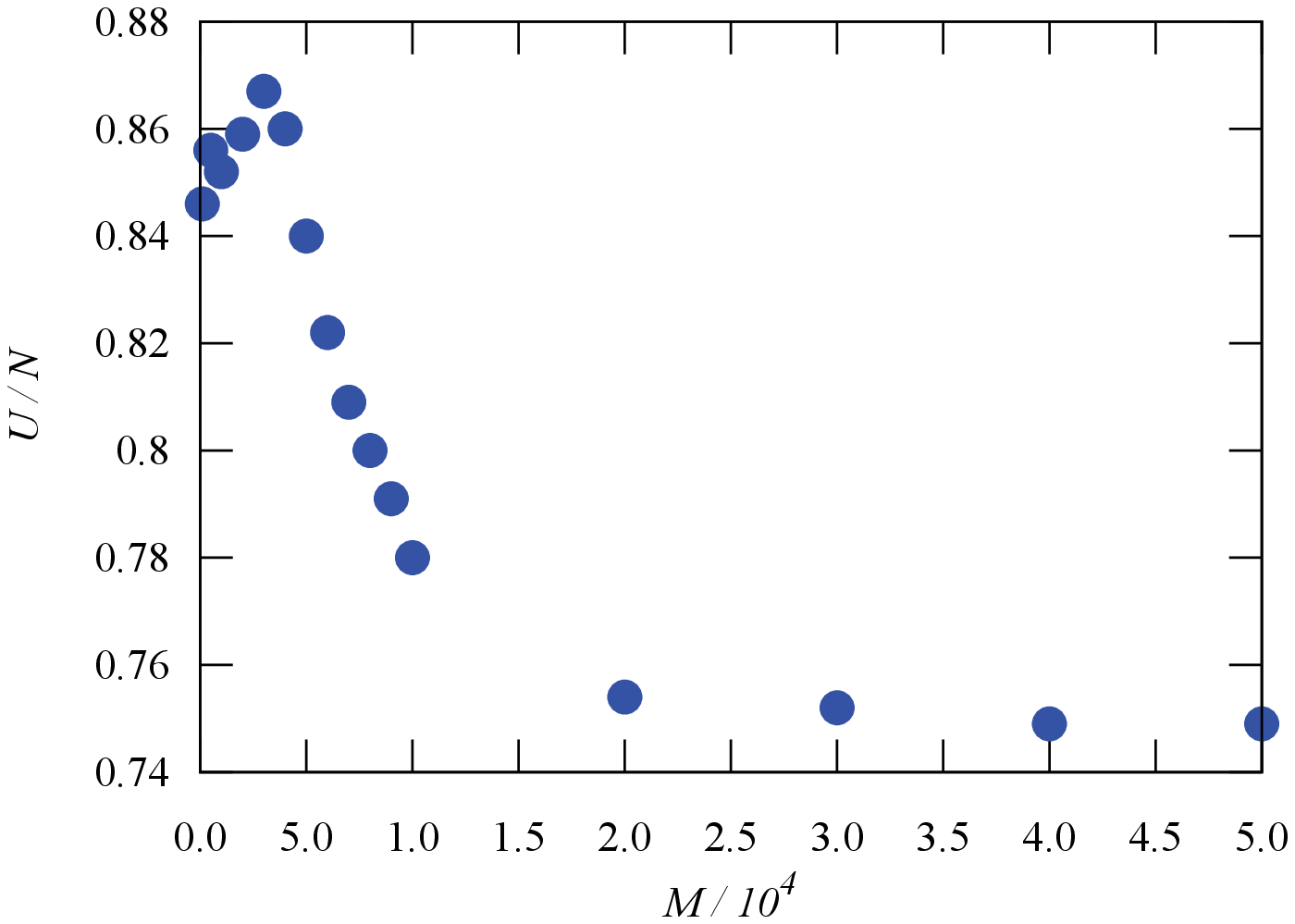}
	\caption{The trial number $M$ dependence on per-agent utility values $U/N$, {for the case of $Q=25$, $H=100$, $T=0.01$, $r=1$, $\bar{\alpha} = 0$, and $\rho_0=0.3$, in which segregation should occur in the correct equilibrium state.}}
	\label{fig:mdep}
\end{figure}

{
	We first examine the impact of trial number $M$ on the per-agent utility values $U/N$ at one time period, for the case of $Q=25$, $H=100$, $T=0.01$, $r=1$, $\bar{\alpha} = 0$, and $\rho_0=0.3$. In this case the segregation should occur in the equilibrium state, i.e., the utility is reduced compared to the initial state
	if the equilibrium state is correctly achieved with sufficiently large value of $M$.
	In Fig.~\ref{fig:mdep}, $U / N$ certainly tends to decrease as $M$ increases: $M > 20000$ is sufficiently large for obtaining the equilibrium state, showing the validity of the value ($M=50000$) we adopt. }

\begin{figure*}[t]
	\centering
	\includegraphics[width=14cm]{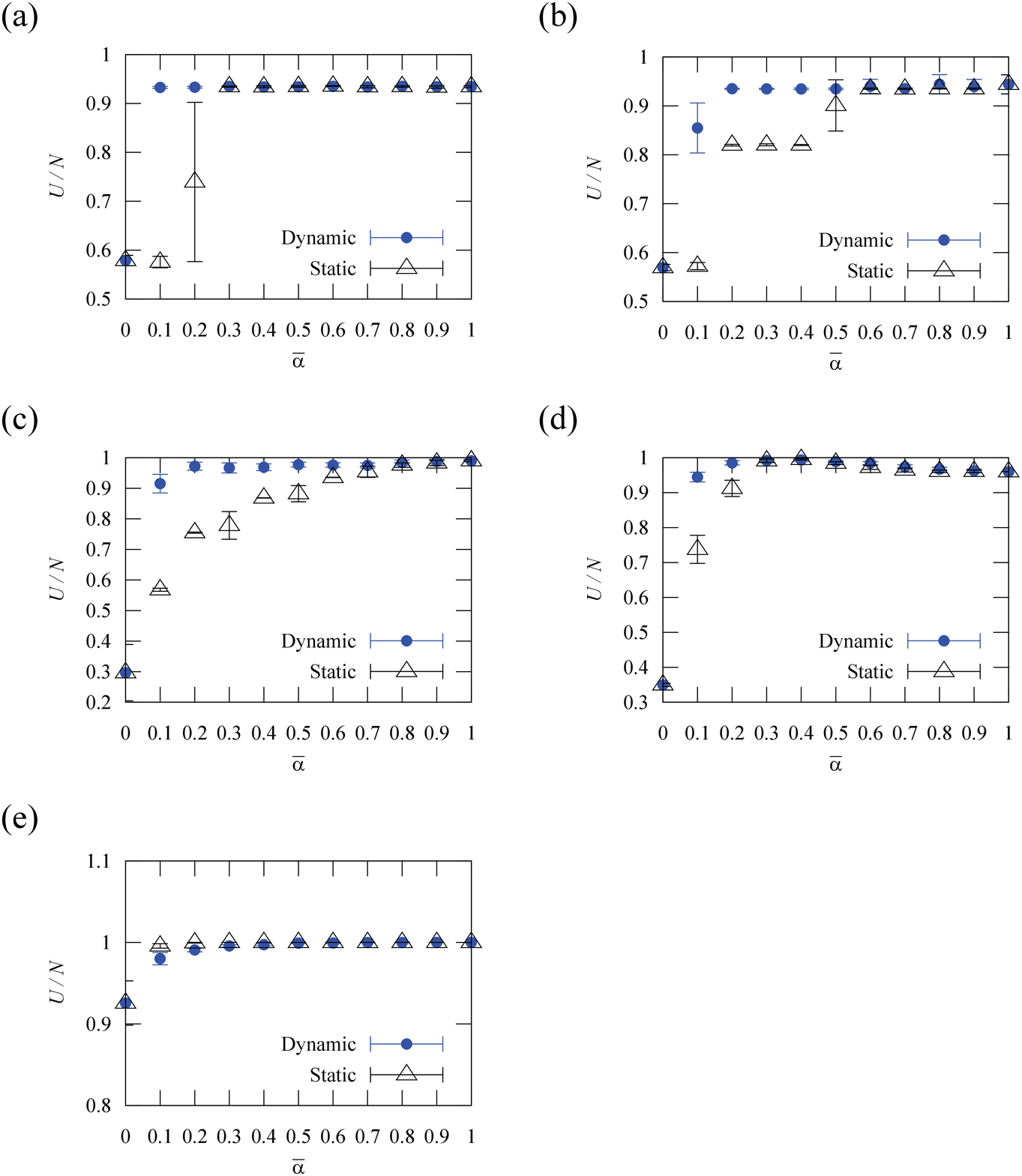}
	\caption{ {Utility per agent $U/N$ as a function of (average) cooperation degree $\bar{\alpha}$
			for (a) $\rho_0 =0.1$, (b) $\rho_0 =0.2$, (c) $\rho_0 =0.3$, (d) $\rho_0 =0.4$, and (e) $\rho_0 =0.5$. 
			The results for static $\bar{\alpha}$ (triangles) and dynamic $\bar{\alpha}$ (filled circles) are shown.} {The other parameters are $Q=25$, $H=100$, $T=0.01$, and $r=0.1$.}
		The symbols indicate the average values for ten simulation runs with different random seeds, and the error bars indicate the variance.}
	\label{fig:utility_compare}
\end{figure*}

Next, we show in Fig.~\ref{fig:utility_compare} the effect of $\bar{\alpha}$ on the per-agent utility values $U/N$ in 
the equilibrium state for various values of $\rho_0$.
The utility is generally low for small values of $\bar{\alpha}$ and tends to increase with $\bar{\alpha}$. 
This is fully consistent with the observation reported in a previous study~\cite{Grauwin2009},
in which the undesired Nash equilibrium for selfish agents and the relaxation thereof due to 
cooperative agents were reported.
In other words, when the value of $\bar{\alpha}$ is small, priority is placed on increasing the individual utility, and thus the system falls in a state split into sparse and dense blocks.
Then, the utility is obviously lower than in a state with more uniformly distributed agents. 
On the other hand, with the large values of $\bar{\alpha}$,
the agents tend to choose to move accompanying the utility increase of the entire city,
resulting in blocks with near-optimal densities around $0.5$.
The density distributions for different values of $\bar{\alpha}$ will be discussed later in more detail.  

The stepwise change in utility is found in Figs.~\ref{fig:utility_compare}(a) ($\rho_0=0.1$)
and \ref{fig:utility_compare}(b) ($\rho_0=0.2$), implying that the bifurcation of states with respect to the parameter.
A drastic change of state is caused, especially for small average densities, such as $\rho_0=0.1$ and $\rho_0=0.2$,
because the impact of the change in the number of blocks with near-optimal densities is large
for such a situation with a small average density.
Accordingly, the variance indicated by the error bars is large around the bifurcation points,
as shown in Fig.~\ref{fig:utility_compare}(a) around $\bar{\alpha}=0.2$,
and Fig.~\ref{fig:utility_compare}(b) around $\bar{\alpha}=0.1$ and $0.5$.

Another important point shown in Fig.~\ref{fig:utility_compare} is that,
in the case of the present model with dynamical cooperative degree,
the utility switches to a large value with smaller values of cooperative degree $\bar{\alpha}$ 
than in the case of a static (constant) cooperative degree.
As expected, the agents with high utility (approximately $\rho=0.5$) accept moves with an increase in the entire utility, even if it sacrifices the individual utility.
This is an effective elimination of the undesired Nash equilibrium with low utility.
A comparison between the static and dynamical cooperation degree models will be
made later herein more comprehensively. 

\begin{figure*}[t]
	\centering
	\includegraphics[width=8cm]{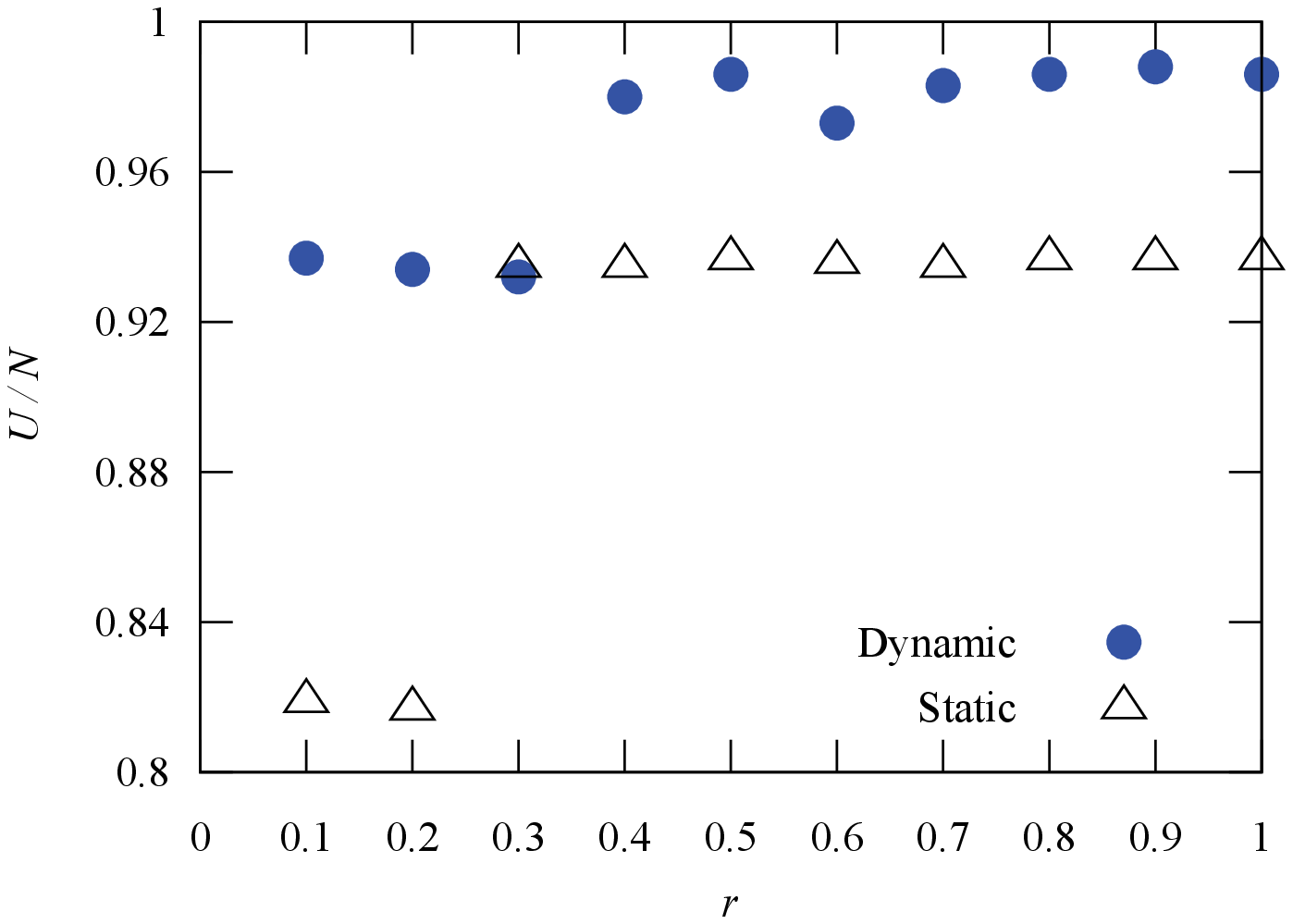}
	\caption{The effect of moving rate $r$ on per-agent utility $U / N$, {for the case of $Q=25$, $H=100$, $T=0.01$, $\rho_0 = 0.2$, and $\bar{\alpha} = 0.5$.}}
	\label{fig:mrate}
\end{figure*}

{Figure~\ref{fig:mrate} plots the utility as a function of the ratio $r$ of moving agents. Overall the impact of increasing $r$ is relatively small, i.e., the utility takes almost the same values for different values of $r$,
	while step-wise increases are observed for $r=0.4$ for the dynamical cooperation degree model and $r=0.3$ for the static cooperation degree model.
}

\begin{figure*}[t]
	\centering
	\includegraphics[width=14cm]{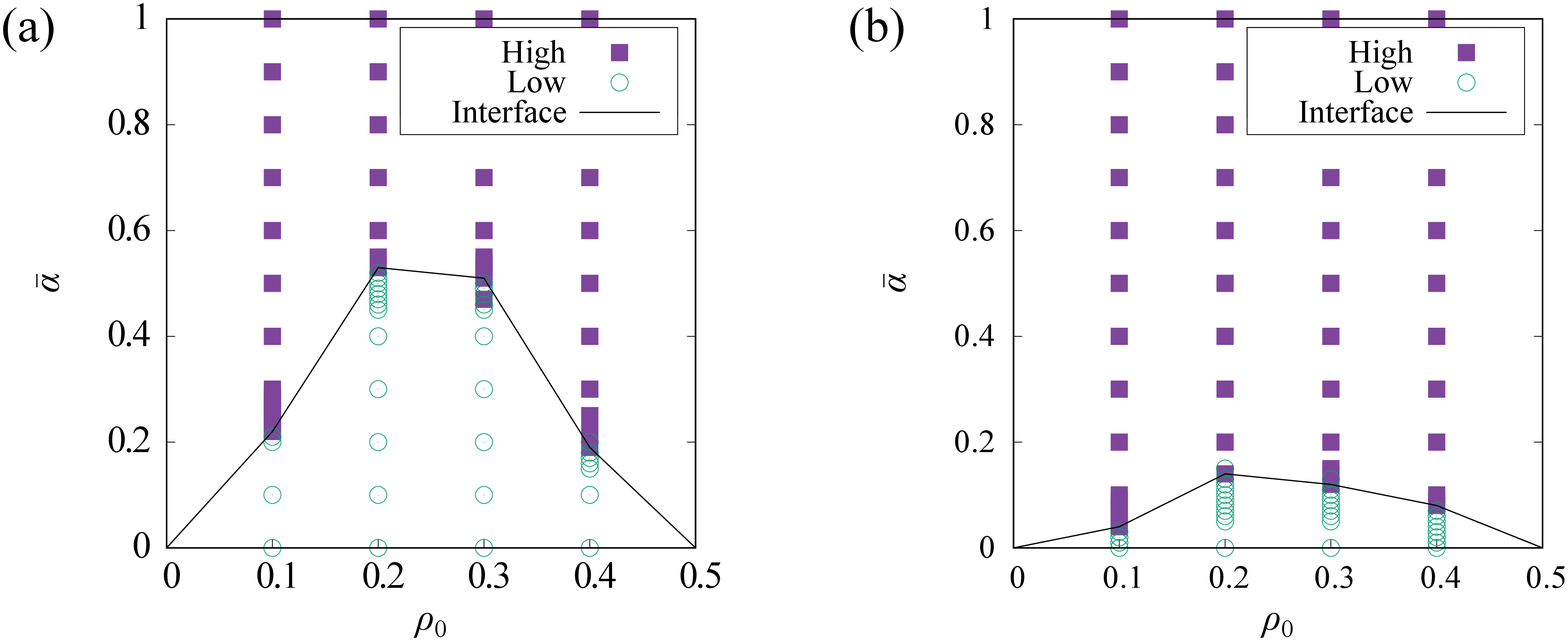}
	\caption{Phase diagram in parameters $\bar{\alpha}$ and $\rho_0$ for (a) the static model and (b) the dynamical model. The moving rate is fixed as $r=0.1$.  {The other parameters are $Q=25$, $H=100$, $T=0.01$.}
		The high-utility phase is defined as the region in which the equilibrium utility exceeds $0.9$.}
	\label{fig:phases}
\end{figure*}

In order to directly compare the efficiency in eliminating the undesired Nash equilibrium,
a phase diagram in the parameter space spanned by $\rho_0$ and $\bar{\alpha}$ is shown in Fig.~\ref{fig:phases}.
More specifically, we define the criteria of the utility value at $0.9$, splitting the phase into 
low-utility and high-utility states, and we obtain the critical value of $\bar{\alpha}$ at which the phase changes. 
As discussed based on Fig.~\ref{fig:utility_compare}, the high cooperative degree values result in
high utility states.
In Fig.~\ref{fig:phases}, the area of high-utility phase is obviously larger for the dynamical cooperation degree model than 
that for the static cooperation degree model. 
This clearly demonstrates the efficient elimination of the undesired Nash equilibrium by means of the dynamical cooperation degree.
In Fig.~\ref{fig:utility_compare}, at the points near the phase boundary, e.g.,
at $\bar{\alpha}=0.2$ for $\rho_0=0.1$ (Fig.~\ref{fig:utility_compare}(a)), and at $\bar{\alpha}=0.5$ for $\rho_0=0.2$ (Fig.~\ref{fig:utility_compare}(b)) in the static cooperative model,
large variances are observed, reflecting the phase change, as indicated by the error bars.

\begin{figure*}[t]
	\centering
	\includegraphics[width=7cm]{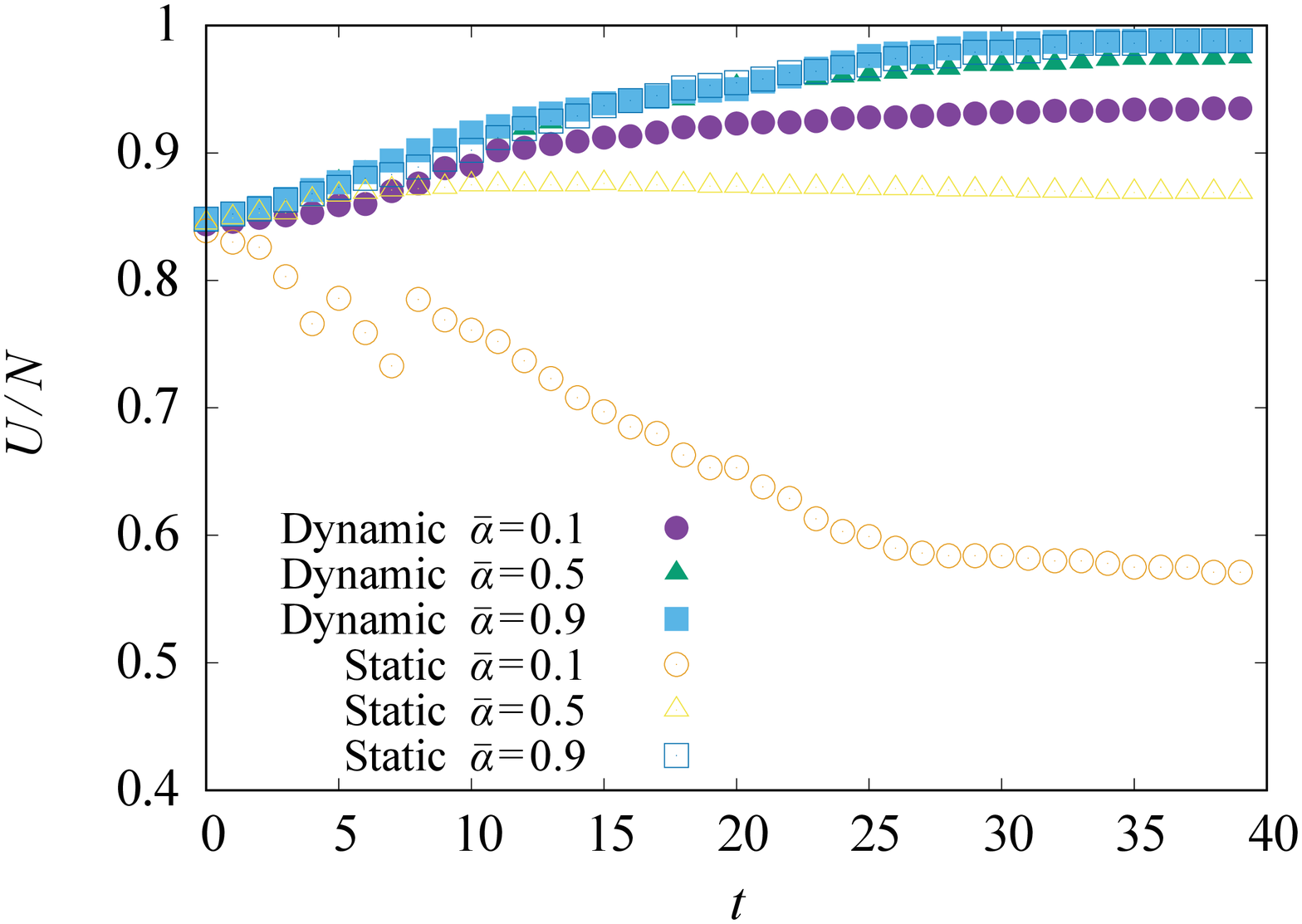}
	\vskip 0.5cm
	\caption{Time dependence of utility per agent $U/N$ for $\rho_0=0.3$ and $r=0.1$. {The other parameters are $Q=25$, $H=100$, $T=0.01$.}}
	\label{fig:Utility_time}
\end{figure*}

We next discuss the time dependence of the utility value.
Figure~\ref {fig:Utility_time} shows the case of $\rho_0 = 0.3$ and $r = 0.1$ (for a certain number of random seeds).
In the case of static cooperation,
for $\bar{\alpha} = 0.1$, the utility decreases with time, indicating the time-dependent process of falling into the undesired Nash equilibrium.
In contrast, for $\bar{\alpha} = 0.5$, the utility increases with time and saturates at the maximum value of $0.869$.
According to Fig.~\ref{fig:utility_compare} (c), in the large gap between $\bar{\alpha}=0.1$ and $\bar{\alpha}=0.5$,
the utility values gradually increase with increasing $\bar{\alpha}$ from $0.1$ to $0.5$.
As shown in Fig.~\ref {fig:phases}, the boundary at which the utility exceeds $0.9$ with the static cooperation model is $\bar{\alpha} = 0.51$ for $r = 0.1$ and $\rho_0 = 0.3$. 
The case of $\bar{\alpha} = 0.5$ is in the low-utility phase, and so is close to the boundary.
At $\bar{\alpha}=0.9$, i.e., well inside the high-utility phase, the utility eventually increases up to approximately unity.
In the case of the dynamical cooperation model, the utility increases to $0.935$,
even when the low value of the cooperation degree is $\bar{\alpha} = 0.1$.
(The phase boundary in Fig.~\ref{fig:phases} for this case is at $\bar{\alpha} = 0.12$ because Fig.~\ref{fig:phases}
is obtained by averaging several simulation runs with different random seed numbers.)
The utility for cases of $\bar{\alpha} = 0.5$ and $\bar{\alpha} = 0.9$ reaches approximately unity.
These results, in particular the comparison between the static and dynamical cooperation degree models for $\bar{\alpha}=0.1$,
demonstrate that even a small (average) value of cooperation degree
has the ability to avoid the undesired Nash equilibrium states in the dynamical cooperation degree model.

\subsection{Distribution of agents}
\begin{figure*}[t]
	\centering
	\includegraphics[width=9cm]{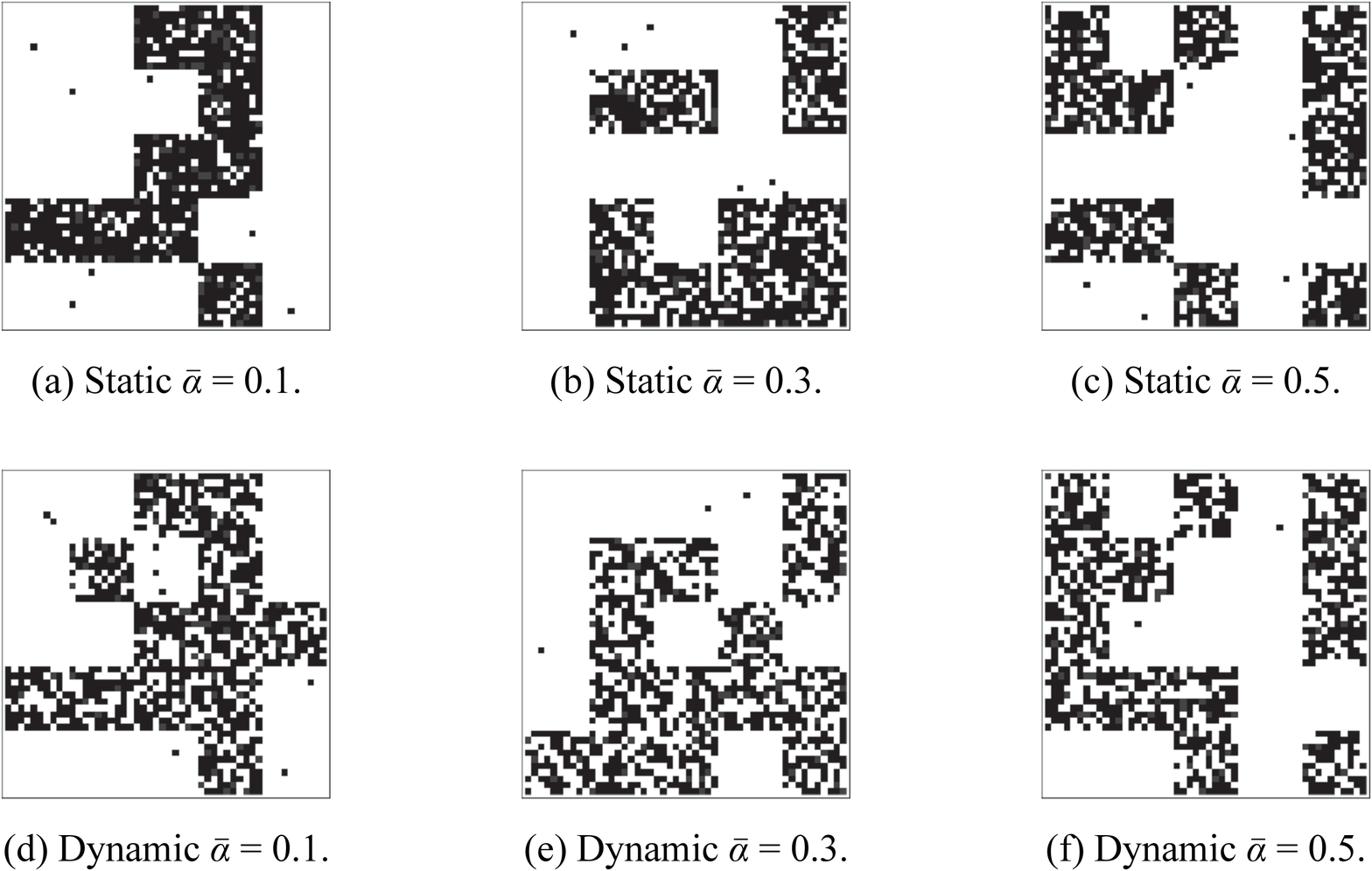}
	\caption{Equilibrium distribution (at $t=40$) of agents for the static and dynamical cooperation degree models for various values of $\bar{\alpha}$ (average value for the dynamical model).
		{The other parameters are $Q=25$, $H=100$, and $T=0.01$.}}
	\label{fig:dist_compare}
\end{figure*}

\begin{figure*}[t]
	\centering
	\includegraphics[width=14cm]{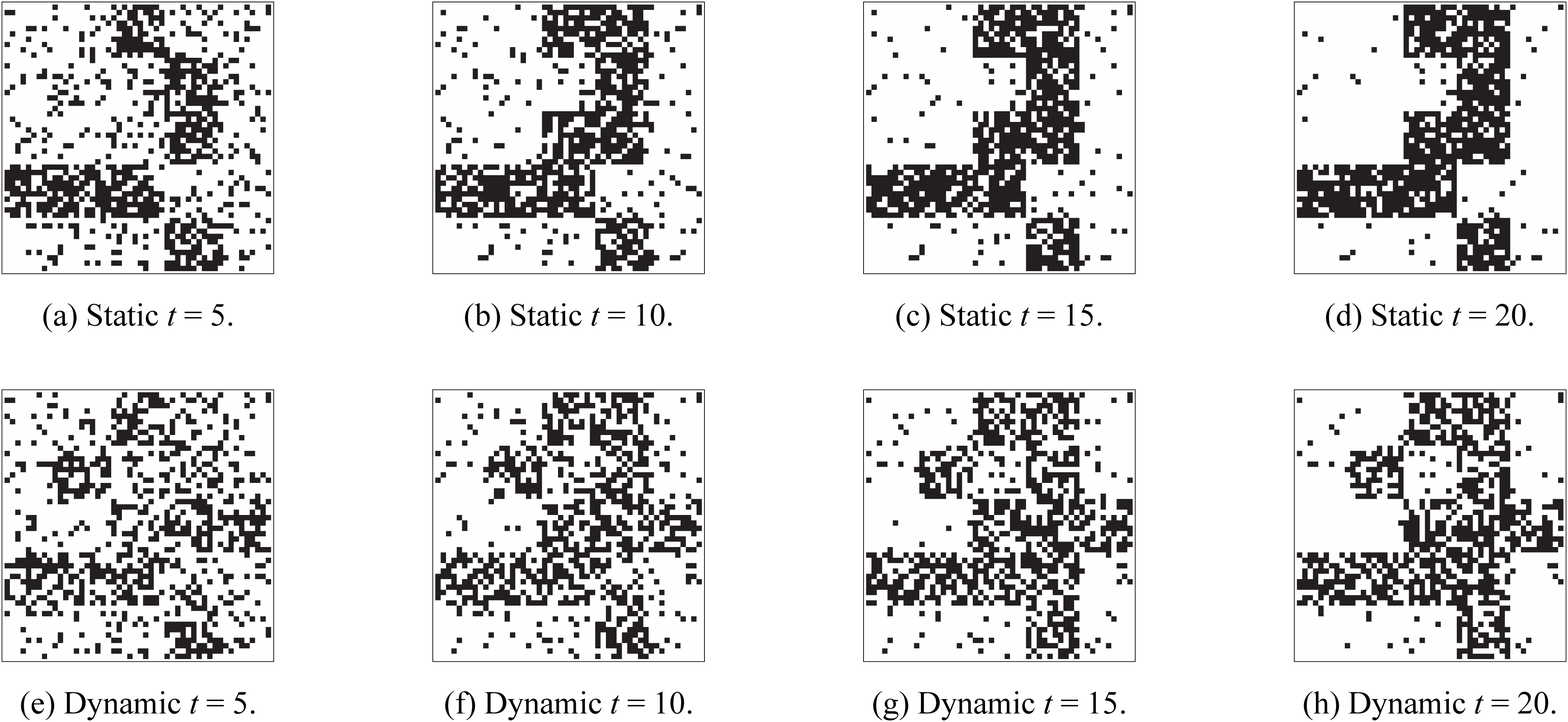}
	\caption{Time evolution of agent distributions for the static model ((a) through (d)) and the dynamical model ((e) through (h)).
		The parameters are $r=0.1$, $\rho_0=0.3$, {$\bar{\alpha}=0.1$, $Q=25$, $H=100$, and $T=0.01$.}}
	\label{fig:dist_time}
\end{figure*}

In order to visually illustrate the effect of the cooperation degree $\bar{\alpha}$ on the distribution of agents in the entire city, in Fig.~\ref{fig:dist_compare}, we show the agent distribution at $t=40$ for $\rho_0 = 0.3$, and the migration rate is $r = 0.1$ (for a certain number of random seeds). Here, the occupied cell ($x^q_i = 1$) is shown in black, 
and vacant cells ($x^q_i = 0$) are shown in white.
For both the static and dynamical cooperation degree models, 
as $\bar{\alpha}$ increases, the number of occupied blocks increases in order to avoid overcrowded states.
For the dynamical model, the number of occupied blocks is larger than that for the static model,
and a non-overcrowded state is realized with small values of $\bar{\alpha}$.
In particular for $\bar{\alpha} = 0.1$, 
although an overcrowded Nash equilibrium is realized with the static model,
the number of blocks occupied increases by three in the dynamical case, which relaxes the density distribution.
In this case, the maximum density of occupied blocks is $0.83$ for the static cooperation model
shown in Fig.~\ref{fig:dist_compare}(a) and $0.64$ for the dynamical cooperation model shown Fig.~\ref{fig:dist_compare}(d).
Increasing $\bar{\alpha}$ as $0.1$, $0.3$, and $0.5$, the maximum density for the static cooperation model varies as $0.83$, $0.68$, and $0.68$, and that for the dynamical cooperation model varies as $0.63$, $0.54$, and $0.58$, respectively.
This result further demonstrates the effective elimination of the overcrowded states by means of the dynamical cooperation model.
In Fig.~\ref{fig:dist_time}, we show the time evolution of the distribution of the agents for the cases of 
$\rho_0=0.3$, $r=0.1$, and $\bar{\alpha}=0.1$ (corresponding to Figs.~\ref{fig:dist_compare}(a) and \ref{fig:dist_compare}(d)).
With the static cooperation model, the density $\rho$ stays between $0.65$ and $0.67$ at $t=5$ in seven blocks,
and the system is about to fall into states in which the density exceeds the maximum utility.
On the other hand, the density $\rho$ for the dynamical cooperation model is $0.54$ at maximum at $t=5$,
showing that the agents indeed gather, but only to the extent that they are not overcrowded.
The maximum density at $t = 20$ is $0.79$ and $0.59$ for the static and dynamical cooperation models, respectively.
The results confirm the effect of the dynamical cooperation degree in the non-equilibrium process.

\clearpage

In summary, the success of the dynamical cooperation degree model in increasing the utility is explained as follows.
With no cooperation ($\bar{\alpha} = 0$), the agents who are initially spread throughout the city gather up to $\rho = 0.5$, which yields the maximum utility.
From low-density blocks, the agents continuously flow into these blocks, resulting in drops of the utility value beyond the maximum value.
This is a one-way flow because flowing back to lower density blocks always decreases the individual utility values. Eventually, a state segregated into overcrowded dense blocks and sparse blocks is established.
On the other hand, with the dynamical cooperation being applied, the cooperation degree is high around $\rho = 0.5$, and thus from blocks having a high density exceeding  $\rho = 0.5$, agents who have the opportunity to move would leave these blocks in order to improve the utility of neighboring agents.
The latter is the driving force to prevent overcrowding.
In addition, in the dynamical coordination degree model, the value of $\alpha$ becomes low in very-high- and very-low-density regions.
Therefore, the move from such blocks should increase the individual utility 
and, at the same time, should increase the degree of cooperation, which further contributes to preventing overcrowding states.
This process is well illustrated in Fig.~\ref {fig:dist_time}, where the accumulation of agents progresses while maintaining the blocks around $\rho = 0.5$.
In the static cooperation degree model, the value of $\alpha$ is not that large around $\rho=0.5$, and hence the effect of preventing the deterioration of utility is weak.
Overall, the present dynamical model
efficiently distributes the cooperation degree to the agents in high-utility blocks, to encourage these agents to move in order to realize uniform states.

\section{Conclusion}\label{sec:conclusion}
In the present study, we have extended the cooperation degree model in Ref.~\cite{Grauwin2009} for Schelling-type segregation models, in order to mitigate the utility declines due to the appearance of undesired Nash equilibriums.
We proposed the dynamical cooperation degree model, 
and investigated the effect on the total utility in the entire city.
More specifically, we used the utility as a quadratic function of local density that takes the maximum value at a density equal to $0.5$ and also defined the cooperation degree parameter as a function of local density. 
The density at which the cooperation degree takes the maximum value is the same as that of the utility function.
With these functions, we implemented a time-dependent numerical procedure that regularly updates the distribution of cooperation degree parameters.
The equilibrium utility values obtained with the present model are compared with
those obtained using the original static cooperation degree model, under the condition that the average value of the cooperation degree is common. 
As a result of our analysis, the dynamical degree of cooperation $\bar{\alpha}$ was found to clearly increase the utility, even for the parameter range in which the previous static degree of cooperation led to undesired low-utility Nash equilibriums.
For example, in the case of $\rho_0=0.3$, the critical value of the cooperation degree 
was $76$\% lower than the static model as shown in Fig.~\ref{fig:phases},
meaning that a small degree of dynamical coordination has the ability to drastically increase the utility of the entire city.

Since the cooperation degree represents {the magnitude with which the Pigouvian tax is imposed}, as pointed out in the previous study \cite{Grauwin2009}, the application of a dynamical tax policy would help control the distribution of residents in the city. In the future, we hope to examine various shapes of the utility function and the coordination function.
Since the non-equilibrium processes via metastable states are affected by utility functions, 
there is a possibility that more effective operation would be achieved depending on the relation between these functions. 
{On the other hand, the costs required for a complex taxation policy should also be taken into account, although estimating the cost for taxation policy is beyond the scope of the present study. Nevertheless developmental calculations, such as discretizing the tax steps to estimate the cost required to actual operation, could be one of the topics of our future work.}

Extending the model city to more realistic ones is also included in our next research topics; for example, as uniform $H$ is simplification of a city, making $H$ non-uniform in order to consider more realistic situations is one possibility of such extensions.

\begin{acknowledgments}
This research was partially supported by Intelligent Mobility Society Design, Social Cooperation Program (Next Generation Artificial Intelligence Research Center, the University of Tokyo, and Toyota Central R\&D Labs., Inc.)
\end{acknowledgments}

\bibliographystyle{apsrev4-1}
\bibliography{arxivDCSchelling}

\end{document}